\documentclass[12pt]{article}
\usepackage{amsmath,amssymb}
\usepackage[dvips]{graphicx}

\newcommand{\bbibitem}[1]{\bibitem{#1}\marginpar{#1}}

\def\Label#1{\label{#1}%
  \smash{\hbox to0pt{\raise1ex\hbox{\tiny[#1]}\hss}}}
\def\noLabels{\let\Label=\label}
\def\nobbibitem{\let\bbibitem=\bibitem}

\newcommand{\be}{\begin{equation}}
\newcommand{\ee}{\end{equation}}
\newcommand{\bea}{\begin{eqnarray}}
\newcommand{\eea}{\end{eqnarray}}
\newcommand{\nn}{\nonumber}
\newcommand{\eq}[1]{(\ref{#1})}
\newcommand{\inv}[1]{\frac{1}{#1}}
\newcommand{\morder}[1]{{\cal O}\left(#1\right) }
\newcommand{\pat}{\partial}

\newcommand{\nmax}{{n_{max}}}
\def\eps{\epsilon}

\begin{document}
%\noLabels
%\nobbibitem

%\rightline{HIP-2007-xx/TH}
%title
%\rightline{arXiv:YYMM.NNNN}
\rightline{HIP-2007-73/TH}
%\leftline{Nov 21, 11:40}
\vskip 1cm \centerline{\large {\bf The %Canonical
Partition Function %of a Multi-Component Coulomb Gas
}}
\vskip 3mm \centerline{\large {\bf %The Canonical Partition Function
of a Multi-Component Coulomb Gas on a Circle}} \vskip 1cm
\renewcommand{\thefootnote}{\fnsymbol{footnote}}
\centerline{{\bf Niko Jokela,$^{1}$\footnote{niko.jokela@helsinki.fi} Matti
J\"arvinen,$^{2}$\footnote{mjarvine@ifk.sdu.dk}%}} \centerline{{\bf
and Esko Keski-Vakkuri$^{1,3}$\footnote{esko.keski-vakkuri@helsinki.fi} }}
\vskip .5cm \centerline{\it ${}^{1}$Helsinki Institute of Physics
and ${}^{3}$Department of Physical Sciences } \centerline{\it
P.O.Box 64, FIN-00014 University of Helsinki, Finland} \centerline{\it ${}^{2}$ University of Southern Denmark, Campusvej 55, DK-5230 Odense M, Denmark}

\setcounter{footnote}{0}
\renewcommand{\thefootnote}{\arabic{footnote}}

\begin{abstract}
%In this note we analyze the partition function of a multicomponent Dyson gas at a special temperature...
We study a two-dimensional Coulomb gas consisting of a mixture of
particles carrying various positive multiple integer charges,
confined on a unit circle. We consider the system in the canonical
and grand canonical ensembles, and attempt to calculate the
partition functions analytically, using Toeplitz and confluent
Vandermonde determinants. Just like in the simple one-component
system (Dyson gas), the partition functions simplify at special
temperature $\beta =2$, allowing us to find compact expressions for
them.
\end{abstract}

\newpage

\section{Introduction and Summary}

Classical two-dimensional Coulomb gas at finite temperature is a prototype example for
learning about phase transitions. There are various interesting simplified
versions of the system. One can confine the
particles to move on a unit circle and yet obtain
an important toy model for various analytical
techniques, which also turns out to be related to other rather different
physical systems \cite{Saleur:1998hq}. 
Typically one restricts all charges to have 
the same strength (which can be
normalized to unity, leaving the temperature as the only parameter)
but possibly different signs. Such systems have been solved exactly in the bulk 
and on the circle 
in the whole stability range of
temperatures \cite{Forrester2,Fendley:1994ms,samaj-2000-101,samaj-2001-105}\footnote{Note also
that a two-component system of charge ratio 1:2 was
considered in \cite{Forrester}.}.
In the simplest system (the Dyson gas \cite{Dyson:1962es}) all the charges are confined
on the circle and have the same sign and strength.
Even the Dyson gas at finite
temperature  is very interesting; among other things
it gives a physical realization of different random matrix
ensembles, just by adjusting the temperature \cite{mehta}. The
canonical partition function can be calculated analytically at
generic temperature and has a simple form \cite{Dyson:1962es, goodwilson}, but the grand canonical partition
function is much more complicated \cite{Saleur:1998hq,Fendley:1994ms,Forrester2}. However, at special inverse temperature
$\beta =2$ it simplifies drastically into a very simple form.

In this paper we consider a slightly more general version of the system,
consisting of a mixture of particles carrying different integer
charges, albeit with the same (positive) sign\footnote{Equivalently, one
could consider a mixture of Dyson gases, at different (initial) temperatures.}.
This generalization
makes the calculations much more complicated. At special inverse
temperature $\beta =2$ there are again simplifications, but now
even the canonical partition function is more difficult to
calculate. We develop techniques using confluent Vandermonde
determinants, Toeplitz determinants and generating functional
methods, to arrive at a compact expression for the partition
function (at $\beta =2$) for any combination of positive integer
charges\footnote{An explicit analytic form can then be found for
any given charge expression with the help of a straightforward
computer algorithm.}. For the grand canonical partition function,
we also find a compact expression, but it is more complicated as it still contains
an infinite series for which we have not been able to find an explicit summation.

Our analysis is mostly restricted to the special inverse temperature $\beta=2$
and to positive integer charges. It would be important to find generalizations of
our techniques for all temperatures. This would also allow a treatment of the case where
some of the particles carry negative
charges \cite{samaj-2000-101,samaj-2001-105,Fendley:1994ms,Jokela:2007wi}.
Another restriction of the present analysis is the number of particles. Although our
formulas are exact, they can in practice only be used for small numbers of multiply
charged particles due to the rapidly increasing complexity of the expressions.
It is perhaps possible to overcome this restriction; instead of exact formulae one may
seek asymptotic approximations for a large number of particles,  e.g. with the help of
asymptotics of large Toeplitz determinants\footnote{We will discuss the behavior of the
partition function in the limit of a large number of $+1$ charges, in a different
framework (decaying branes in string theory) in a future publication \cite{NMEinprep}.}.
Together with a generalization to arbitrary temperatures, this might shed light on the
thermodynamic limit of these systems, including phase transitions and analytic properties
of the partition functions.

The article is organized as follows.
We introduce the multi-component
Coulomb gas in the next section. In section 3, we construct
the system from the one-component Dyson gas by placing several unit charges
at the same point and removing the infinite self-energy, and present a result
for the canonical partition function of the multi-component gas. Technical details of the calculation
involving manipulations of confluent Vandermonde determinants are postponed to appendix \ref{appA}.
In section 4 we develop the former construction into a generating functional method
to compute the canonical  and grand canonical partition functions.
We end by a giving an explicit example of applying our methods in section 5.

\section{Setup}

The Hamiltonian of the standard Dyson gas with $N$ particles is given by
\be \label{standham}
 H_\mathrm{D} = \sum_{1\le i<j}^N V(t_i,t_j) \ ,
\ee
where
\be
 V(t_i,t_j) = - \log|e^{it_i}-e^{it_j}|
\ee
is the two-dimensional Coulombic potential between the particles $i$ and $j$ which are located at $e^{it_i}$ and at $e^{it_j}$ on the unit circle, respectively.
The definition can be naturally extended for particles having multiple positive integer charges.
Let us take $N_1$ particles of charge $+1$, $N_2$ particles of charge $+2$, \ldots , $N_\nmax$ particles of charge $+\nmax$, labeled by positions $\{t^{(1)}_{1},\ldots , t^{(1)}_{N_1}\}$ , $\{t^{(2)}_1,\ldots,t^{(2)}_{N_2}\}$, \ldots, $\{t^{(\nmax)}_1,\ldots, t^{(\nmax)}_{N_\nmax}\}$, respectively. They interact via
\be
 V_{nm}(t_i^{(n)},t_j^{(m)}) = -nm\log|e^{it_i^{(n)}}-e^{it_j^{(m)}}| \ .
\ee
The Hamiltonian of this system is then given by 
\bea
 H &=& \sum_\mathrm{pairs} V_{nm}(t_i^{(n)},t_j^{(m)}) \\\nn
 &=&-\sum_{n=1}^\nmax n^2\sum_{1\leq i<j}^{N_n}\log|e^{it_i^{(n)}}-e^{it_j^{(n)}}|-\sum_{1\leq n<m}^\nmax nm\sum_{i=1}^{N_n}\sum_{j=1}^{N_m}\log|e^{it_i^{(n)}}-e^{it_j^{(m)}}| \ .
\eea
The canonical partition function is defined by
\be \label{ZCdef}
 Z_{\rm{C}}(N_1,N_2,\ldots,N_\nmax) = \frac{1}{\prod_n N_n!}\int\left[\prod_{n=1}^\nmax\prod_{i=1}^{N_n}\frac{dt_i^{(n)}}{2\pi}\right] e^{-\beta H} %=  \frac{I(N_1,N_2,N_3,\ldots)}{\prod_n N_n!}
  \ .
\ee
In this paper we analyze $Z_{\rm C}$ at a fixed inverse temperature $\beta=2$. The results are also used to calculate the grand canonical partition function\footnote{For an attempt to calculate the grand canonical partition function of this system, but in a different framework, see \cite{Jokela:2007dq}.}
\be \label{ZGdef}
 Z_{\rm G}(\hat z_1,\ldots,\hat z_\nmax) = \sum_{N_1=0}^\infty (\hat z_1)^{N_1}\cdots \sum_{N_\nmax=0}^\infty (\hat z_\nmax)^{N_\nmax} Z_{\rm C}(N_1,\ldots,N_\nmax) \ ,
\ee
where $\hat z_i$ are the fugacities which correspond to the different charges. We will start by a rather direct evaluation of the  integrals over $t_i$ in \eq{ZCdef}, and then go on to study a more advanced formulation involving generating functionals.

%at $\beta=2$ is doable. Here it goes.

\section{$Z_{\rm C}$ and confluent Vandermonde determinants}
\label{VanSec}

Let us study the partition function \eq{ZCdef} at the particular inverse temperature $\beta=2$. %the canonical partition function simplifies. In particular,
We denote $z_i=\exp(it_i)$. The partition function of the standard Dyson gas with the Hamiltonian \eq{standham} becomes
\be \label{DysonZ}
 Z_{\rm D}(N) = \inv{N!}\int \prod_{i=1}^N \frac{dt_i}{2\pi} \prod_{1\le i<j}^N\left|e^{it_i}-e^{it_j}\right|^2 = \inv{N!}\oint \prod_{i=1}^N \frac{dz_i}{2\pi iz_i} \left|\Delta(z_1,\ldots,z_N)\right|^2 \ .
\ee
Here the integrand $\rho_{\rm D} = \exp\left(-2H_{\rm D}\right)$ is the absolute value squared of the Vandermonde determinant
\bea \label{Ddef}
|\Delta(z_1,\ldots,z_N)|^2 &=& \prod_{1\leq i<j\leq N}|z_i-z_j|^2 =\left|\sum_{\{i\}}\varepsilon_{i_1 \cdots i_N} z_1^{i_1-1} \cdots z_N^{i_N-1}\right|^2 \nn \\
&=& \left|\sum_\Pi (-1)^{\Pi} \prod_{i=1}^N z_i^{\Pi(i)-1}\right|^2 \ ,
\eea
where $\Pi$ denotes permutations of ${1,2,\ldots,N}$ and $(-1)^{\Pi}$ is the sign of the permutation $\Pi$. %Above $|z_i|=1$ was used in the last step.
Using $|\Delta|^2=\Delta \Delta^*$ we can express the integrand as an analytic function of $z_i$,
\be \label{delsq}
 |\Delta(z_1,\ldots,z_N)|^2 = \sum_{\Pi_1,\Pi_2}(-1)^{\Pi_1}(-1)^{\Pi_2} \prod_{i=1}^N z_i^{\Pi_1(i)-\Pi_2(i)} \ ,
\ee
and doing the integrals in \eq{DysonZ} simply picks the constant term in \eq{delsq} such that $Z_{\rm D}(N) =1$.

The above analysis generalizes to the multi-component Coulomb gas. In fact, %it is well known [] that
\be
\prod_{\rm pairs}(z_i^{(n)}-z_j^{(m)})^{nm}
\ee
can be expressed as a determinant of a confluent Vandermonde matrix (see, \emph{e.g.}, \cite{Kalman}).
We present here some details of a simple proof since in the case of the Dyson gas it has an intuitive physical interpretation. The general treatment can be found in appendix A.

\begin{figure}[ht]
\begin{center}
\noindent
\includegraphics[width=0.8\textwidth]{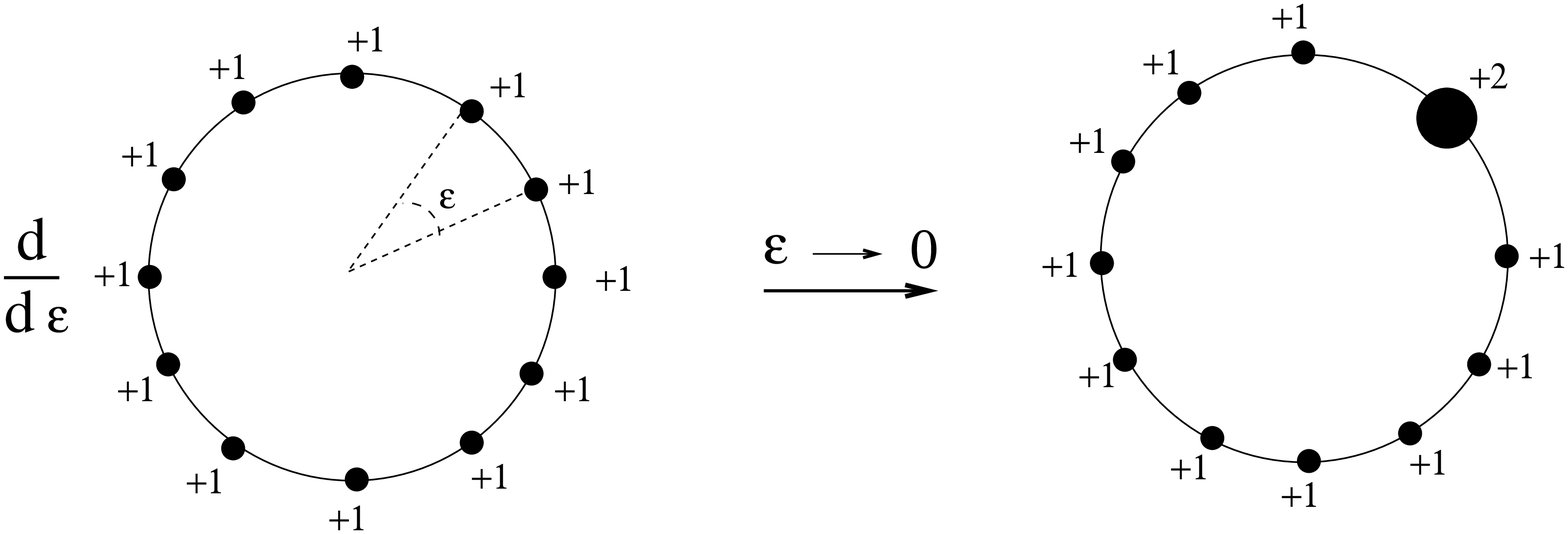}
\end{center}
\caption{(The square root of) the Boltzmann weight of a state with one doubly-charged particle is obtained by forcing two unit charges to coincide. The infinite self-energy of the formed pair is removed by a derivative wrt. the angle $\eps$ between the particles.}
\label{epsfig}
\end{figure}

The idea of the derivation goes as follows. We start from a single-component Dyson gas with at a lot of particles. We then form multiply-charged particles by placing several unit charges at the same point and removing the infinite self-energy of the formed particle.
Let us start from the simplest case of only one double charge to see how this works (see also figure \ref{epsfig}). We take a system with $N+2$ unit charges and denote $\tau_i=t_{N+i}$, $w_i=z_{N+i}$. Setting $\tau_2 = \tau_1+\eps$ we immediately get
\bea
 |\Delta(z_1,\ldots,z_{N+2})| &=&
 \prod_{1\leq i<j\leq N}|z_i-z_j|\prod_{i=1}^N|z_i-w_1||z_i-w_2||w_1-w_2| \nn\\
 &=& \prod_{1\leq i<j\leq N}|z_i-z_j|\prod_{i=1}^N|z_i-w_1|^2\eps + \morder{\eps^2} \nn\\
  &\equiv& \left[\rho_2(z_1,\ldots,z_N,w_1)\right]^{1/2}\eps+ \morder{\eps^2} \ ,
\eea
where the leading result gives exactly the wanted Boltzmann weight $\rho_2$ times the exponential of the self-energy that we need to divide away. Notably, as depicted in figure \ref{epsfig} the above procedure  equals differentiation,
\bea \label{rhores}
 \left[\rho_2(z_1,\ldots,z_N,w_1)\right]^{1/2}&=&\inv{\eps} |\Delta(z_1,\ldots,z_{N+2})|+ \morder{\eps} \nn\\
 &=& \left|\frac{\Delta(z_1,\ldots,z_N,w_1,w_2)}{w_2-w_1}\right|+ \morder{\eps} \nn\\
 &\substack{\longrightarrow \\ \eps\to 0}&  \left|\frac{\partial}{\partial w_2}\Delta(z_1,\ldots,z_N,w_1,w_2)\right|_{w_2=w_1} \\\nn
 &=& \left|\sum_\Pi (-1)^{\Pi} \prod_{i=1}^N z_i^{\Pi(i)-1} w_1^{\Pi(N+1)-1} [\Pi(N+2)\!-\!1] w_1^{\Pi(N+2)\!-\!2}\right|
\eea
where \eq{Ddef} was used in the last step. In particular, from \eq{rhores} one sees that $\rho$ can be expressed as a determinant of a certain confluent Vandermonde matrix,
\be
 A = \left(\begin{array}{cccccc}
            1   & 1   & \cdots   & 1   & 1   & 0 \\
        z_1 & z_2 &          & z_N & w_1 & 1 \\
        z_1^2 & z_2^2 &      & z_N^2 & w_1^2 & 2 w_1 \\
        \vdots &&&&& \vdots \\
        z_1^{N+1} & z_2^{N+1} & \cdots & z_N^{N+1} & w_1^{N+1} & (N+1)w_1^N
       \end{array} \right)
\ee
which is obtained from the standard Vandermonde matrix by differentiation.  A lengthy calculation gives the formula
\bea \label{onedoubres}
 Z_{\rm C}(N,1) &=& \inv{N!1!}\int \prod_{i=1}^{N} \frac{dt_i}{2\pi} \frac{d\tau}{2\pi} \rho_2(t_1,\ldots,t_N,\tau) \nn\\
 &=& \inv{N!}\oint \prod_{i=1}^N \frac{dz_i}{2\pi iz_i} \frac{dw_1}{2\pi iw_1}\sum_{\Pi_1,\Pi_2} (-1)^{\Pi_1} (-1)^{\Pi_2} \prod_{i=1}^{N} z_i^{\Pi_1(i)-\Pi_2(i)}\nn\\
 &&\times [\Pi_1(N+2)-1][\Pi_2(N+2)-1] w_1^{\Pi_1(N+1)+\Pi_1(N+2)-\Pi_2(N+1)-\Pi_2(N+2)} \nn\\
 &=& \inv{12} (N+3)(N+2)^2(N+1) = {N+4\choose 4} + {N+3\choose 4}  \ ,
\eea
where the complex integrals again pick the constant term of the integrand.
The result \eq{onedoubres} equals the one obtained by using the Selberg integral \cite{Jokela:2007dq}.

We present the straightforward, but technical generalization to any number of charges in the appendix~\ref{appA}. In particular, particles carrying higher charges ($n>2$) are obtained by taking higher derivatives. The final result for $Z_{\rm C}$ in a general configuration is given in \eq{resgen}. Notice that the result can be generalized to some special higher values of $\beta$ by adjusting the charge configuration of the particles. This is possible if $\sqrt{\beta/2} q_i$ are positive integers for all the charges $q_i$ of the system. For example, $Z_{\rm C}$ for a system with $\beta=8$ and for $\tilde N_n$ particles of the charge $n$ (with $n=1,\ldots ,\, \nmax$)  is obtained by setting
\bea
 N_1=N_3=\cdots =0\ \mathrm{;} \qquad N_{2n} = \tilde N_n
\eea
in \eq{resgen}.

Similar results for $Z_{\rm C}$ as those of the appendix~\ref{appA} can also be derived by using a more elegant method that involves a generating functional, which we will discuss next.

\section{Generating functional method} \label{fdersec}

Let us define a generating functional (see appendix \ref{appB} for an alternative definition of $I[j]$),
\be \label{Idef}
 I[j] = \sum_{M=0}^\infty \inv{M!} \int \left[\prod_{i=1}^M \frac{dt_i}{2\pi}j(t_i)\right] \prod_{1\le i <j}^M \left|e^{it_i}-e^{it_j}\right|^\beta \ .
\ee
The weight of the standard Dyson gas $\rho_{\rm D}$ is then found by taking functional derivatives of $I[j]$,
\be \label{dysden}
 \rho_{\rm D}(t_1,\ldots,t_N) = \left.\frac{\delta^N I[j]}{\delta j(t_1) \cdots \delta j(t_N)}\right|_{j=0} \ .
\ee
For the special inverse temperature $\beta=2$ we can use \eq{delsq} to write (see, \emph{e.g.}, \cite{Balasubramanian:2004fz})
\bea \label{Iasdet}
I[j] &=& \sum_{M=0}^\infty \inv{M!} \int \left[\prod_{i=1}^M\frac{dt_i}{2\pi}j(t_i)\right]\sum_{\Pi_1,\Pi_2}(-1)^{\Pi_1}(-1)^{\Pi_2} \prod_{i=1}^M e^{it_i[\Pi_1(i)-\Pi_2(i)]} \nn\\
&=& \sum_{M=0}^\infty \inv{M!} \sum_{\Pi_1,\Pi_2}(-1)^{\Pi_1}(-1)^{\Pi_2} \prod_{i=1}^M \hat j_{\Pi_2(i)-\Pi_1(i)} \ ,
\eea
where $\Pi_i$ are permutations of the length $M$ and
\be
 \hat j_n = \int \frac{dt}{2\pi} j(t) e^{-int}
\ee
are the Fourier coefficients of the source. That is, $I[j]$ is a sum over the Toeplitz determinants of $\hat j$,
\be \label{Zres}
 I[j] = \sum_{M=0}^\infty \det T^{(M)}[j] \ ,
\ee
where $T^{(M)}$ is an $M \times M$ matrix with the entries
\be \label{Tdef}
 \left(T^{(M)}[j]\right)_{kl} = \hat j_{k-l}\ .
\ee

One might expect that the formula \eq{dysden} can be generalized to particles with higher charges by taking higher functional derivatives at a single point, or letting some of the $t_i$'s in \eq{dysden} coincide. However, as above in section \ref{VanSec} this leads to a problem with the infinite self-energy. Luckily, as above in \eq{rhores}, the self-energy can be removed by taking suitable derivatives wrt. $t_i$. For example, the weight for a state with $N$ unit charges and one double charge can be expressed as
\bea \label{normres}
 \rho_2(t_1,\ldots,t_N,\tau) &=& \rho_{\rm D}(t_1,\ldots,t_N,\tau,\tau+\eps)\eps^{-2}  + \morder{\eps} \nn\\
 &=&
 \left. \inv{2!} \left(\frac{\pat}{\pat \tau_2}\right)^2\rho_{\rm D}(t_1,\ldots,t_N,\tau_1,\tau_2)\right|_{\tau=\tau_1=\tau_2} \nn\\
 &=&
\left. \inv{2!} \left(\frac{\pat}{\pat \tau_2}\right)^2\frac{\delta^{N+2} I[j]}{\delta j(t_1) \cdots \delta j(t_N) \delta j(\tau_1) \delta j(\tau_2)}\right|_{\substack{j=0 \\ \tau=\tau_1=\tau_2}} \ ,
\eea
where $\eps^{-2} = \exp\left(2 E_{\rm self}\right)$ was added to remove the self-energy.
As seen from \eq{normres} combining the differentiation idea of the previous section \eq{rhores} to the functional derivative result \eq{dysden} leads to a ``normalized'' second order functional derivative at a single point,
\be \label{D2def}
 D_2 \sim e^{+2 E_{\rm self}} \int \frac{dt}{2\pi} \frac{\delta^2}{\delta^2 j(t) } = \int \frac{d\tau}{2\pi} \left[\inv{2!} \left(\frac{\pat}{\pat \tau_2}\right)^2 \frac{\delta^2}{\delta j(\tau_1) \delta j(\tau_2)}\right]  _{\tau=\tau_1=\tau_2} \ ,
\ee
%after the subtraction of the infinite self-energy.  and expressed in terms of the Fourier modes $\hat j_n$ of the source $j(t)$.
where the infinite self-energy has been removed.

We shall now work out a generalization of \eq{D2def} for $D_n$ with $n>2$. It can be conveniently expressed in terms of the Fourier modes $\hat j_n$ of the source $j(t)$ [see the definition \eq{Dndef} below]. Let us consider the case where one particle (at $e^{i\tau}$) carries an arbitrary integer charge $n$, which can be added by using the result \eq{derres} in appendix \ref{appA}.
For $n>2$ \eq{normres} generalizes to
%The weight of the state $\rho_n$ is given by
\bea \label{wder}
 \rho_n(t_1,\ldots,t_N,\tau) &=& \left.\prod_{k=2}^n \inv{(2k-2)!}\left(\frac{\pat}{\pat \tau_k}\right)^{2k-2}\rho_{\rm D}(t_1,\ldots,t_N,\tau_1,\ldots,\tau_n)\right|_{\tau=\tau_1=\cdots=\tau_n} \\\nn
 &=&\left.\frac{(-1)^{n(n-1)/2}}{n!\left[(n-1)!\right]^n}\prod_{k=1}^n \left(\frac{\pat}{\pat \tau_k}\right)^{n-1}\!\!\rho_{\rm D}(t_1,\ldots,t_N,\tau_1,\ldots,\tau_n)\right|_{\tau=\tau_1=\cdots=\tau_n} .
\eea
In particular, the definition in terms of derivatives can be written in various forms. Any combination is fine as long as the sum of the orders of the derivatives is $n(n-1)$ and the normalization factor $K$ of the appendix \ref{appA} can be worked out. We shall use the latter form of \eq{wder}. Then the partition function reads
\bea
 Z_{{\rm C},\, n}(N) &=& Z_{\rm C}(N_1=N,N_2=0,\ldots,N_{n-1}=0,N_n=1) \\\nn
&=& \frac{(-1)^{n(n-1)/2}}{N!\,n!\left[(n-1)!\right]^n}\int \prod_{i=1}^N \frac{dt_i}{2\pi}\frac{d\tau}{2\pi} \\\nn
 &&\left\{ \left(\frac{\pat}{\pat \tau_1}\right)^{n-1}\!\!\cdots\ \left(\frac{\pat}{\pat \tau_n}\right)^{n-1}\!\!\frac{\delta^{N+n}I[j]}{\delta j(t_1) \cdots \delta j(t_N)\delta j(\tau_1) \cdots \delta j(\tau_n)}\right\}_{\substack{j=0\\ \tau=\tau_1=\cdots=\tau_n}} \! .
\eea
The partition function can now be evaluated by using the result \eq{Zres}. The chain rule implies
\be
 \frac{\delta}{\delta j(t)} = \sum_{m=-\infty}^\infty \frac{\delta \hat j_m}{\delta j(t)}\frac{\pat}{\pat\hat j_m} = \sum_{m=-\infty}^\infty e^{-imt}\frac{\pat}{\pat \hat j_m} \ .
\ee
Only the terms that include products of exactly $N+n$ of $\hat j_m$'s give non-zero contribution which fixes $M=N+n$ in \eq{Zres}. Hence
\bea \label{onepres}
 Z_{{\rm C},\, n} &=&  \frac{(-1)^{n(n-1)/2}}{N!\,n!\left[(n-1)!\right]^n} \int \frac{d\tau}{2\pi}  \left\{  \left(\frac{\pat}{\pat \tau_1}\right)^{n-1}\!\!\cdots\ \left(\frac{\pat}{\pat \tau_n}\right)^{n-1}\!\! \right.\\
  &&\times\left. \sum_{m_1=-M}^M \!\!\!\cdots\!\!\!\sum_{m_n=-M}^M\! \exp\!\left[-i\sum_{k=1}^n m_k \tau_k\right]\left(\frac{\pat}{\pat \hat j_0}\right)^N \prod_{k=1}^n \frac{\pat}{\pat \hat j_{m_k}} T^{(N+n)}[j]\right\}_{\substack{j=0\\ \tau=\tau_1=\cdots=\tau_n}}\nn\\\nn
  &=&\left.\inv{N!}\left(\frac{\pat}{\pat \hat j_0}\right)^N\!\!\! \sum_{m_1=-M}^M \!\!\!\cdots\!\!\!\sum_{m_n=-M}^M\!\! \frac{\delta\left(\sum_{k=1}^n m_k,0\right)}{n!\left[(n-1)!\right]^n}\!\left[\prod_{k=1}^n (m_k)^{n\!-\!1} \frac{\pat}{\pat \hat j_{m_k}}\right]\det  T^{(M)}[j]\right|_{j=0}
\eea
where $\delta(m,n)$ denotes the Kronecker delta. Here the operator\footnote{Note that the $D_n$ are not unique. For example, using the first line in \eq{wder} would lead to $D_n = \sum_{m_1,\ldots,m_n=-M}^M \delta_{\sum_{k=1}^n m_k,0}\prod_{k=1}^n \frac{(-m_k^2)^{k-1}}{(2k-2)!} \frac{\pat}{\pat \hat j_{m_k}}$. }
\be \label{Dndef}
 D_n = \sum_{m_1=-M}^M \!\!\!\cdots\!\!\!\sum_{m_n=-M}^M\!\frac{\delta\left(\sum_{k=1}^n m_k,0\right)}{n!\left[(n-1)!\right]^n}\prod_{k=1}^n (m_k)^{n-1} \frac{\pat}{\pat \hat j_{m_k}}
 %D_n = \sum_{m_1,\ldots,m_n=-M}^M \delta_{\sum_{k=1}^n m_k,0}\prod_{k=1}^n \frac{(-m_k^2)^{k-1}}{(2k-2)!} \frac{\pat}{\pat \hat j_{m_k}} \ .
\ee
generates the particle with the charge $n$ in the gas. The result \eq{Dndef} is valid for $\beta=2$, but depends on temperature, in general. The temperature dependence arises from the self-energies which can be removed by differentiation similarly as above at all integer-valued inverse temperatures. Thus \eq{Dndef} can be easily generalized to all $\beta=1,2,\ldots$. However, the generating functional can be written in terms of Toeplitz determinants as in \eq{Iasdet} only for $\beta=2$.

Generalization of \eq{onepres} to any number of multiply-charged particles is straightforward. If we set $M=\sum_{n=1}^\nmax n N_n$ in the definition \eq{Dndef}, then\footnote{Notice that $D_1 =\frac{\pat}{\pat \hat j_0}$.}
\be \label{ZCres}
 Z_{\rm C}(N_1,\ldots,N_\nmax) = \left.\left[\prod_{n=1}^\nmax \frac{\left(D_n\right)^{N_n}}{N_n!} \right]\det T^{(M)} [j]\right|_{j=0} \ .
\ee
Inserting the definition \eq{Tdef} one can express $Z_{\rm C}$ as a finite sum similarly as in appendix~\ref{appA} (see also the concrete example below).

Also, a compact formal expression for the grand canonical partition function \eq{ZGdef} at $\beta=2$ immediately follows,
\be\label{eq:ZG}
 Z_{\rm G}(\hat z_1,\ldots,\hat z_\nmax)
 =  \left.\exp\left[\sum_{n=1}^\nmax \hat z_n D_n\right] \sum_{M=0}^\infty \det T^{(M)} [j]\right|_{j=0} \ ,
\ee
where the summation limits in the definition \eq{Dndef} need to be extended from $\pm M$ to $\pm\infty$.

\section{An example}

%We shall generalize the formula \eq{onepres}  to the multicomponent gas below, but before doing that we
As a concrete example of the use of \eq{ZCres} continue from \eq{onepres} and show how a purely combinatorial result is derived in the case of only one multiply-charged particle. By inserting the explicit form of $\det T^{(M)}[j]$ from \eq{Iasdet} we find
\bea \label{ZCnres}
 Z_{{\rm C},\,n} &=& \left.\inv{N!} D_1^N D_n\det T^{(M)} [j]\right|_{j=0} \nn\\
 &=& \frac{1}{N!\, {n!\left[(n-1)!\right]^n}}\sum_{m_1=-M}^M \!\!\!\cdots\!\!\!\sum_{m_n=-M}^M\!\delta\left(\sum_{k=1}^n m_k,0\right) \prod_{k=1}^n (m_k)^{n-1} \nn\\
 &&\times\sum_{\Pi_1,\Pi_2}(-1)^{\Pi_1}(-1)^{\Pi_2} \prod_{i=1}^{N} \delta_{\Pi_1(i),\Pi_2(i)}\prod_{j=1}^n \delta_{\Pi_1(j+n)+m_j,\Pi_2(j+n)} \ .
\eea
In this special case of only one multiply-charged particle, the condition $\sum_{k=1}^n m_k = 0$ follows the others in \eq{ZCnres}, and may be dropped. The conditions $\Pi_1(i)=\Pi_2(i)$ for $i=1,\ldots,N$ define a certain ``subdeterminant'', which may be written as
\bea \label{oneinsres}
 Z_{{\rm C},\,n} &=& \frac{1}{\left[(n-1)!\right]^n}\!\!\! \sum_{1\le \ell_1< \cdots<\ell_n}^M\!\! \sum_{\tilde \Pi}(-1)^{\tilde \Pi} \prod_{k=1}^n \left[\ell_k-\ell_{\tilde \Pi(k)}\right]^{n-1} \ ,
\eea
where $\tilde \Pi$ is a permutation of length $n$.

%\newpage

\bigskip
\bigskip

\noindent
{\bf \large Acknowledgments}

\bigskip

N.J. thanks the hospitality of the University of Southern Denmark during the preparation of this article. N.J. has been in part supported by the Magnus Ehrnrooth foundation. M.J. has been in part supported by the Magnus Ehrnrooth foundation and by the Marie Curie Excellence Grant under contract MEXT-CT-2004-013510. This work was also partially supported by the EU 6th Framework
Marie Curie Research and Training network ``UniverseNet'' (MRTN-CT-2006-035863).

\bigskip

\bigskip

\appendix

\section{A general combinatorical formula for $Z_{\rm C}$} \label{appA}

In this appendix we derive a general expression for $Z_{\rm C}$ using confluent Vandermonde determinants and compute the consequent combinatorical result.  For the general multi-component Coulomb gas, \eq{Ddef} reads
\be \label{Vanres}
 \prod_\mathrm{pairs}\left|z_i^{(n)}-z_j^{(m)}\right|^{2nm} = |\det A|^2 \ ,
\ee
where $A$ is a $M \times M$ confluent Vandermonde matrix with $M=\sum_{n=1}^\nmax n N_n$. $A$ is defined by
\be \label{Adef}
 A = \left(A^{(1)} \left(z_1^{(1)}\right) \cdots A^{(1)}\left(z_{N_1}^{(1)}\right) %\ A^{(2)}[z_1^{(2)}] \cdots A^{(2)}[z_{N_2}^{(2)}]
 \cdots A^{(\nmax)}\left(z_1^{(\nmax)}\right) \cdots A^{(\nmax)}\left(z_{N_\nmax}^{(\nmax)}\right) \right) \ ,
\ee
where each $ M \times n$ subblock $A^{(n)}$ corresponds to one particle of the gas and has the entries
\be \label{Asubdef}
 \left(A^{(n)}(z)\right)_{ij} = \frac{1}{(j-1)!}\left(\frac{\partial}{\partial z}\right)^{j-1} z^{i-1} =
 \left\{ \begin{array}{rcl}
             {i-1 \choose j-1} z^{i-j} & \mathrm{if} & i \ge j \\
             0                         & \mathrm{if} & i < j
         \end{array} \right.
\ee

Equation \eq{Vanres} with the definition \eq{Adef}, \eq{Asubdef} for $A$ is a known result, but for use elsewhere we will sketch how it follows from the standard Vandermonde determinant \eq{Ddef}. Let $f(t_1,\ldots ,t_N,\tau_1,\ldots,\tau_n)$ be a function such that
\be \label{funcdef}
 f(t_1,\ldots,t_N,\tau_1,\ldots,\tau_n) = g(t_1,\ldots,t_N,\tau_1,\ldots,\tau_n) \prod_{1\le i < j}^n (\tau_i-\tau_j)^{k_{ij}} \ ,
\ee
where $k_{ij}$ are positive integers and $g$ stays finite and non-zero as any $\tau_i \to \tau_j$. Then take the $m_i$th derivative wrt. $\tau_i$ for each $i=1,2,\ldots,n$ of the both sides of \eq{funcdef} such that $m_i\ge 0$ and
\be
 \sum_{i=1}^n m_i =  \sum_{1\le i < j}^n k_{ij} \ ,
\ee
and let $\tau=\tau_1=\cdots=\tau_n$ in the end. On the right hand side only such terms survive, where all the derivatives act on the product. Thus
\be \label{derres}
 g(t_1,\ldots,t_N,\tau,\ldots,\tau) = \inv{K} \left.\left(\frac{\pat}{\pat \tau_1}\right)^{m_1}\!\!\!\cdots\left(\frac{\pat}{\pat \tau_n}\right)^{m_n} f(t_1,\ldots,t_N,\tau_1,\ldots,\tau_n)\right|_{\tau=\tau_1=\cdots=\tau_n} \ ,
\ee
where $K$ is the constant integer\footnote{The result only makes sense if $K\ne 0$.}
\be
 K = \left(\frac{\pat}{\pat \tau_1}\right)^{m_1}\!\!\!\cdots\left(\frac{\pat}{\pat \tau_n}\right)^{m_n}\prod_{1\le i < j}^n (\tau_i-\tau_j)^{k_{ij}} = \tilde K \prod_{k=1}^n m_k!
\ee
where $\tilde K$ is the coefficient of the monomial $\tau_1^{m_1}\cdots\tau_n^{m_n}$ in $\prod_{1\le i < j}^n (\tau_i-\tau_j)^{k_{ij}}$.
To prove \eq{Vanres} for the case of $N$ unit charges and one particle with charge $n$, we apply the result \eq{derres} to
\bea
 f(z_1,\ldots,z_N,w_1,\ldots,w_n) &=& \Delta(z_1,\ldots,z_N,w_1,\ldots,w_n) \nn\\
   &=& \prod_{1\le i<j}^n(w_i-w_j) \times \prod_{1\le i<j}^N (z_i-z_j) \prod_{i=1}^N \prod_{j=1}^n(z_i-w_j) \nn\\
   &=& \prod_{1\le i<j}^n(w_i-w_j) \times g(z_1,\ldots,z_N,w_1,\ldots,w_n)
\eea
and take $m_k=k-1$.
After calculating the integer $K = \prod_{k=1}^n(k-1)!$ we find
\bea
 \prod_{1\le i<j}^N \!(z_i\!-\!z_j) \prod_{i=1}^N(z_i\!-\!w)^n &\!\!=\!\!&  g(z_1,\ldots,z_N,w,\ldots,w)  \nn\\
            &\!\!=\!\!&\left.\prod_{k=1}^n \inv{(k\!-\!1)!}\! \left(\!\frac{\pat}{\pat w_k}\!\right)^{\!\!k-1\!}\!\!\! \Delta(z_1,\ldots,z_N,w_1,\ldots,w_n)\right|_{w=w_1=\cdots =w_n}\nn\\
        &\!\!=\!\!& \pm\det \left(A^{(1)} \left(z_1\right) \cdots A^{(1)}\left(z_{N}\right) A^{(n)}\left(w\right)\right) \ ,
\eea
where we used the determinant form of $\Delta$ in \eq{Ddef} and \eq{Asubdef} in the last line to complete the proof.
% a $N+n \times N+n$ Vandermonde matrix $B$. Then
% \be
%  \det B = \prod_{}()
% \ee
The general result [equation \eq{Vanres} with \eq{Adef}] follows by a straightforward induction.

% The entries of $A$ are
% \be
%  A_{ij} = \frac{1}{(s-1)!}\left(\frac{\partial}{\partial z_{k}^{(n)}}\right)^{s-1} \left(z_k^{(n)}\right)^{j-1}
% \ee
% The relation between $n,k,s$ and $i$ is (uniquely) determined by $1\le n$, $1\le k \le N_n$, $1\le s \le n$ and $\ell(n,k)+s=i$ with $\ell(n,k)=\sum_{m=1}^{n-1} m N_m + (k-1)n$.
In the rest of this appendix, we use the explicit form \eq{Adef}, \eq{Asubdef} of $A$ to find a combinatorical formula for $Z_{\rm C}$, the constant term in all $z_i^{(n)}$ of \eq{Vanres} when expanded into an analytic function of the $z_i^{(n)}$.
Note that since we only need the determinant of $A$ and $|z_i^{(n)}| = 1$ we may multiply any column by an arbitrary power of $z_i^{(n)}$. Using in addition elementary column operations $A^{(n)}(z)$ may be replaced by
\be
 \left(\tilde A^{(n)}(z)\right)_{ij} = \frac{i^{j-1}}{(j-1)!} z^i \ .
\ee
Then we find
\bea
 && |\det A|^2  \nn\\ %\prod_\mathrm{pairs}\left|z_i^{(n)}-z_j^{(m)}\right|^{2nm} \nonumber\\
% &=& \left|\sum_{\{i\}}\varepsilon_{i_1 \cdots i_M} \prod_n \prod_{k=1}^{N_n} \left[\prod_{s=1}^{n}\frac{1}{(s-1)!}\left(\frac{\partial}{\partial z_{k}^{(n)}}\right)^{s-1} \left(z_{k}^{(n)}\right)^{i_{\ell(n,k)+s}-1}\right]  \right|^2 \\
&=& \left|\sum_{\{i\}}\varepsilon_{i_1 \cdots i_M} \prod_{n=1}^\nmax \prod_{k=1}^{N_n} \left[\prod_{s=1}^{n}\frac{\left(i_{\ell(n,k)+s}\right)^{s-1}}{(s-1)!} \left(z_{k}^{(n)}\right)^{i_{\ell(n,k)+s}}\right]  \right|^2 \nonumber\\
&=& \left|\sum_{\{i\}}\varepsilon_{i_1 \cdots i_M} \prod_{n=1}^\nmax \prod_{k=1}^{N_n} \frac{1}{n!(n-1)!\cdots 1!}\Delta(i_{\ell(n,k)+1},\ldots,i_{\ell(n,k)+n}) \left(z_k^{(n)}\right)^{\sum_{s=1}^n i_{\ell(n,k)+s}} \right|^2 \nonumber
\eea
where %$z_k^{(n)}=\exp(it_k^{(n)})$, $|z_k^{(n)}|^2=1$ was used in the last step, and
the Vandermonde matrices of the permutation variables in the last form are obtained after antisymmetrization.
%Note that the %complicated
The function
\be
 \ell(n,k)= \sum_{m=1}^{n-1} m N_m + (k-1)n
\ee
is only needed for picking up the permutation variable $i$ with the correct index.

The constant term of $|\det A|^2 = \det A^* \det A$ is
\bea
 Z_{\rm C}(N_1,\ldots,N_\nmax) \prod_{n=1}^\nmax N_n! \!\! &=&\!\!\sum_{\{i\},\{j\}} \varepsilon_{i_1 \cdots i_M} \varepsilon_{j_1 \cdots j_M}\prod_{n=1}^\nmax \prod_{k=1}^{N_n} \frac{1}{\left[n!(n-1)!\cdots 1!\right]^2}\\
 \!\!&&\!\!\times \Delta(i_{\ell(n,k)+1},\ldots,i_{\ell(n,k)+n})\Delta(j_{\ell(n,k)+1},\ldots,j_{\ell(n,k)+n})\nonumber\\\nonumber
 \!\!&&\!\!\times  \delta\left(i_{\ell(n,k)+1}+\cdots+i_{\ell(n,k)+n},j_{\ell(n,k)+1}+\cdots+j_{\ell(n,k)+n}\right)
\eea
where $\delta(i,j)=\delta_{ij}$ is the Kronecker $\delta$-symbol.
For $n=1$ the $\delta$ restrictions give simply $i_k=j_k$. Using these the result evaluates to
\bea
 Z_{\rm C} \prod_{n=2}^\nmax N_n! %\frac{I(N_1,N_2,\ldots)}{N_1!}
 \!\!&=&\!\! \sum_{S,\{i\},\{j\}} \varepsilon_{i_1 \cdots i_K} \varepsilon_{j_1 \cdots j_K}\prod_{n=2}^\nmax \prod_{k=1}^{N_n} \frac{1}{\left[n!(n-1)!\cdots 1!\right]^2}\\
 \!\!&&\!\!\times \Delta\left(S(i_{\ell'(n,k)+1}),\ldots,S(i_{\ell'(n,k)+n})\right)\Delta\left(S(j_{\ell'(n,k)+1}),\ldots,S(j_{\ell'(n,k)+n})\right)\nonumber\\
 \!\!&&\!\!\times  \delta\left(\sum_{s=1}^n S(i_{\ell'(n,k)+s}),\sum_{s=1}^n S(j_{\ell'(n,k)+s})\right) \ , \nonumber
\eea
where $K=M-N_1$, the first sum goes over all increasing functions $S:\{1,\ldots,K\} \to \{1,\ldots,M\} $ [so that $i<j \Leftrightarrow S(i)<S(j)$], and $\ell'(n,k) = \ell(n,k)-N_1$.

Due to symmetry, one can add the restrictions $i_{\ell'(n,k)+1}<i_{\ell'(n,k+1)+1}$ (for all $n>1$ and $1\le k<N_n$), and
$i_{\ell'(n,k)+s}<i_{\ell'(n,k)+s+1}$, $j_{\ell'(n,k)+s}<j_{\ell'(n,k)+s+1}$ (for all $n>1$, $k$, and $1\le s < n$) and multiply by the ratio of numbers of terms ($\prod_{n=2}^\nmax n!^{N_n} N_n!$). Then the result becomes
\bea \label{resgen}
 Z_{\rm{C}} %\!\!&=&\!\!\frac{I(N_1,N_2,\ldots)}{\prod_n N_n!}\nonumber\\
 \!\!&=&\!\! \sum_{S,\{i\},\{j\}}\!\!\!' \ \varepsilon_{i_1 \cdots i_K} \varepsilon_{j_1 \cdots j_K}\prod_{n=2}^\nmax \prod_{k=1}^{N_n} \frac{1}{\left[(n-1)!\cdots 1!\right]^2}\\
 \!\!&&\!\!\times \Delta\left(S(i_{\ell'(n,k)+1}),\ldots,S(i_{\ell'(n,k)+n})\right)\Delta\left(S(j_{\ell'(n,k)+1}),\ldots,S(j_{\ell'(n,k)+n})\right)\nonumber\\
 \!\!&&\!\!\times  \delta\left(\sum_{s=1}^n S(i_{\ell'(n,k)+s}),\sum_{s=1}^n S(j_{\ell'(n,k)+s})\right) \ , \nonumber
\eea
where the prime indicates the presence of the above restrictions.
% In particular,
% \bea \label{resn2}
%  %\hat I_2(N_1,N_2) &=& \hat I(N_1,N_2,0,0,\ldots)
%  Z_{\rm C}(N_1,N_2) %&=& \nonumber\\
%  \!\!&=&\!\! \sum_{S,\{i\},\{j\}}\!\!\!' \ \varepsilon_{i_1 \cdots i_{K}} \varepsilon_{j_1 \cdots j_{K}} \prod_{k=1}^{N_2}
%   \left(S(i_{2k-1})-S(i_{2k})\right)\left(S(j_{2k-1})-S(j_{2k})\right)\nonumber\\
%  \!\!&&\!\!\times  \delta\left(S(i_{2k-1})+S(i_{2k}),S(j_{2k-1})+S(j_{2k})\right)
% \eea
% where $K=2N_2$ and $\ell'(2,k)=2(k-1)$ was inserted.
%We have written c-codes which evaluate $I$ using the formulae (\ref{resgen}), (\ref{resn2}) for a given (but arbitrary) set of $\{N_n\}$.

\section{Relation to field theory}\label{appB}

In this appendix we present a compact formulation (for all inverse temperatures $\beta$) of the generating functional \eq{Idef} using the language of field theory. We use the operator $X(t)$ that lies at $z=e^{it}$ on the boundary of the unit disk. Its self contraction
\be
 \langle  X(t_1) X(t_2) \rangle = \log\left|e^{it_1}-e^{it_2}\right|^2
\ee
is essentially the Coulomb potential between charges at $e^{it_1}$ and at $e^{it_2}$. Then the generating functional \eq{Idef} reads
\be \label{ICFT}
 I[j] = \left\langle  \exp\left[\int\frac{dt}{2\pi}j(t):e^{\sqrt{\beta/2}\, X(t)}:\right]  \right\rangle
\ee
as can be verified by using the Wick theorem. %In particular,
The canonical %and grand canonical
partition function of the (single-component) Dyson gas reads
\be
 Z_{\rm C,D}(N) =  \inv{N!}\left\langle \left[\int\frac{dt}{2\pi}:\exp\left(\sqrt{\beta/2}\, X(t)\right):\right]^N\right\rangle = \left.\inv{N!}\left[\int \frac{dt}{2\pi} \frac{\delta}{\delta j(t)}\right]^N I[j]\right|_{j=0}
\ee
where the last step can be checked using \eq{ICFT}.
% The grand canonical partition function can also be written in a compact form,
% \be
%  Z_{\rm G,D}(\hat z) =  \left.\left\langle :\exp\left[\hat z \int\frac{dt}{2\pi}e^{\sqrt{\beta/2}X(t)}\right]:\right\rangle = \exp\left[\hat z\int \frac{dt}{2\pi} \frac{\delta}{\delta j(t)}\right]I[j]\right|_{j=0} \ .
% \ee

In the case of the multi-component gas
\be
 Z_{\rm C}(N_1,\ldots,N_\nmax) = \inv{\prod_{n=1}^\nmax (N_n)!}\left\langle\prod_{n=1}^\nmax\left[\int\frac{dt}{2\pi}:e^{n\sqrt{\beta/2}\, X(t)}:\right]^{N_n} \right\rangle
\ee
so that one would naively expect that
\be \label{Znaive}
 Z_{\rm C}(N_1,\ldots,N_\nmax) \sim \left.\inv{\prod_{n=1}^\nmax (N_n)!}\prod_{n=1}^\nmax\left[\int\frac{dt}{2\pi} \frac{\delta^n}{\left(\delta j(t)\right)^n}\right]^{N_n}  I[j]\right|_{j=0} \ .
\ee
However, \eq{Znaive} fails in general since the higher order functional derivatives at the same point are not compatible with the normal ordering: the right hand side includes self contractions between fields at the same point. These are the self-energies of the multiply-charged particles which were discussed in the text. Equation \eq{Znaive} holds only after the subtraction of the self-energies, which was done for $\beta=2$ explicitly in section~\ref{fdersec}, leading to \eq{ZCres}. Notice also that the grand canonical partition function can be written as (see \cite{Jokela:2007dq})
\be
 Z_{\rm G}(\hat z_1,\ldots,\hat z_\nmax) = \left\langle \exp\left[\sum_{n=1}^\nmax \hat z_n \int\frac{dt}{2\pi}:e^{n \sqrt{\beta/2}X(t)}:\right] \right\rangle
\ee
which explains the exponential form of the result \eq{eq:ZG}.


\begin{thebibliography}{99}

%\cite{Saleur:1998hq}
\bibitem{Saleur:1998hq}
  H.~Saleur,
  %``Lectures on non perturbative field theory and quantum impurity  problems,''
  arXiv:cond-mat/9812110.
  %%CITATION = COND-MAT/9812110;%%

%\cite{Forrester2}
\bibitem{Forrester2}
  P.~J.~Forrester,
  %``Solvable isotherms for a two-component system of charged rods on a line,''
 J.\ Statist.\ Phys.\  {\bf 51}, 457 (1988);
  %%CITATION = ??;%%
% P.~J.~Forrester,
  %``Exact results for correlations in a two-component log-gas,''
 J.\ Statist.\ Phys.\  {\bf 59}, 57 (1989).
  %%CITATION = ??;%%

%\cite{Fendley:1994ms}
\bibitem{Fendley:1994ms}
  P.~Fendley, F.~Lesage and H.~Saleur,
  %``Solving 1-d plasmas and 2-d boundary problems using Jack polynomials and
  %functional relations,''
  J.\ Statist.\ Phys.\  {\bf 79}, 799 (1995) [arXiv:hep-th/9409176].
  %%CITATION = HEP-TH 9409176;%%

 \bibitem{samaj-2000-101}
  L.~Samaj and I.~Travenec,
  %``Thermodynamic Properties of the Two-Dimensional Two-Component Plasma,''
  J.\ Stat.\ Phys.\ {\bf 101} (2000) 713 [arXiv:cond-mat/0004021].

  \bibitem{samaj-2001-105}
  L.~Samaj,
  %``Thermodynamic Properties of the One-Dimensional Two-Component Log-Gas,''
   J.\ Stat.\ Phys.\ {\bf 105} (2001) 175 [arXiv:cond-mat/0102332].

%\cite{Forrester}
\bibitem{Forrester}
  P.~J.~Forrester,
  %``Interpretation of an Exactly Solvable Two-Component Plasma,''
  J.\ Statist.\ Phys.\  {\bf 35} 77 (1984).
  %%CITATION = JSTPB,35,77;%%

%\cite{Dyson:1962es}
\bibitem{Dyson:1962es}
  F.~J.~Dyson,
  %``Statistical theory of the energy levels of complex systems. I,''
  J.\ Math.\ Phys.\  {\bf 3} (1962) 140.
  %%CITATION = JMAPA,3,140;%%

\bibitem{mehta}
  M.~L.~Mehta, {\em Random Matrices}, 2nd edition, Academic Press
  (1991).

\bibitem{goodwilson}
 K.~G.~Wilson, %{\em Proof of conjecture by Dyson},
 J.\ Math.\ Phys.\
 {\bf 3} (1962) 1040; I.~J.~Good, %{\em Short proof of a conjecture by Dyson},
 J.\ Math.\ Phys.\ {\bf 11} (1970) 1884.

%\cite{Jokela:2007wi}
\bibitem{Jokela:2007wi}
  N.~Jokela, E.~Keski-Vakkuri and J.~Majumder,
  %``Timelike Boundary Sine-Gordon Theory and Two-Component Plasma,''
  Phys.\ Rev.\  D {\bf 77}, 023523 (2008)
  [arXiv:0709.1318 [hep-th]].
  %%CITATION = PHRVA,D77,023523;%%

\bibitem{NMEinprep}
  N.~Jokela, M.~J\"arvinen and E.~Keski-Vakkuri,
  \emph{in preparation}.
  %%CITATION = ARXIV:0705.1916;%%

%\cite{Jokela:2007dq}
\bibitem{Jokela:2007dq}
  N.~Jokela, M.~J\"arvinen, E.~Keski-Vakkuri and J.~Majumder,
  %``Disk Partition Function and Oscillatory Rolling Tachyons,''
  J. Phys. A: Math. Theor. {\bf 41} (2008) 015402 [arXiv:0705.1916 [hep-th]].
  %%CITATION = ARXIV:0705.1916;%%

%\cite{Sen:2002nu}
%\bibitem{SLNT}
%  A.~Sen,
%  %``Rolling tachyon,''
%  JHEP {\bf 0204}, 048 (2002)
%  [arXiv:hep-th/0203211];
%  %%CITATION = JHEPA,0204,048;%%
%  F.~Larsen, A.~Naqvi and S.~Terashima,
%  %``Rolling tachyons and decaying branes,''
%  JHEP {\bf 0302} (2003) 039
%  [arXiv:hep-th/0212248].
%  %%CITATION = JHEPA,0302,039;%%


\bibitem{Kalman}
  %The Generalized Vandermonde Matrix
  D.~Kalman, Mathematics Magazine, {\bf 57} (1984) 15. %Vol. 57, No. 1. (Jan., 1984), pp. 15-21.
  %Stable URL: http://links.jstor.org/sici?sici=0025-570X%28198401%2957%3A1%3C15%3ATGVM%3E2.0.CO%3B2-9


%\cite{Balasubramanian:2004fz}
\bibitem{Balasubramanian:2004fz}
  V.~Balasubramanian, E.~Keski-Vakkuri, P.~Kraus and A.~Naqvi,
  %``String scattering from decaying branes,''
  Commun.\ Math.\ Phys.\  {\bf 257} (2005) 363
  [arXiv:hep-th/0404039];
  %%CITATION = CMPHA,257,363;%%
  N.~Jokela, E.~Keski-Vakkuri and J.~Majumder,
  %``On superstring disk amplitudes in a rolling tachyon background,''
  Phys.\ Rev.\  D {\bf 73} (2006) 046007
  [arXiv:hep-th/0510205].
  %%CITATION = PHRVA,D73,046007;%%
%
% %\cite{Dyson:1962es}
% \bibitem{BJKM}%Dyson:1962es}
% %  F.~J.~Dyson,
% %  %``Statistical theory of the energy levels of complex systems. I,''
% %  J.\ Math.\ Phys.\  {\bf 3} (1962) 140.
% %  %%CITATION = JMAPA,3,140;%%
%   V.~Balasubramanian, N.~Jokela, E.~Keski-Vakkuri and J.~Majumder,
%   %``A thermodynamic interpretation of time for rolling tachyons,''
%   Phys.\ Rev.\  D {\bf 75} (2007) 063515
%   [arXiv:hep-th/0612090].
%   %%CITATION = PHRVA,D75,063515;%%



\end{thebibliography}
\end{document}